\documentclass[twocolumn,aps,prb,superscriptaddress]{revtex4-1}
\usepackage{graphicx,amsmath}
\usepackage{chapterbib}
\usepackage{dcolumn}
\usepackage{bm}
\usepackage{subfigure}

\begin{document}
\title{Superconductivity near lattice instability: the case of NbC$_{1-x}$N$_x$ and NbN}

\author{Simon Blackburn}
\author{Michel C\^ot\'e}
\affiliation{D\'epartement de physique, Universit\'e de Montr\'eal, C. P. 6128 Succursale Centre-ville, Montr\'eal (Qu\'ebec) H3C 3J7, Canada}
\author{Steven G. Louie}
\author{Marvin L. Cohen}
\affiliation{Department of Physics, University of California at Berkeley and Materials Sciences Division, Lawrence Berkeley National Laboratory, Berkeley, CA 94720}

\begin{abstract}

Using density functional theory within the local density approximation we report a study of the electron-phonon coupling in NbC$_{1-x}$N$_x$ and NbN crystals in the rocksalt structure.  The Fermi surface of these systems exhibits important nesting.  The associated Kohn anomaly greatly increases the electron-phonon coupling and induces a structural instability when the electronic density of states reaches a critical value. Our results reproduce the observed rise in $T_{c}$ from 11.2~K to 17.3~K as the nitrogen doping is increased in NbC$_{1-x}$N$_x$. To further understand the contribution of the structural instability to the rise of the superconducting temperature, we develop a model for the Eliashberg spectral function in which the effect of the unstable phonons is set apart. We show that this model together with the McMillan formula can reproduce the increase of $T_{c}$ near the structural phase transition. 
\end{abstract}
\maketitle

\section{Introduction}

 The discovery of superconductivity in boron-doped diamond renewed interest in phonon-mediated type superconductors.\cite{Ekimov:2004p69} Some of these materials like Cs$_{3}$C$_{60}$~\cite{Ganin:2008p221} and MgB$_{2}$~\cite{Nagamatsu:2001p107} are reported with transition temperature $T_{c}$ up to almost 40~K. In the case of MgB$_{2}$, it has been suggested that the Kohn anomaly is responsible for the enhanced electron-phonon coupling leading to a larger critical temperature $T_{c}$.\cite{PhysRevLett.92.197004} This phenomenon has been argued to raise electron-phonon coupling in copper oxide superconductors.\cite{Giustino:2008p209} In the case of transition metal carbides, the relatively high $T_{c}$ is also explained by a Kohn anomaly.\cite{Noffsinger:2008p93} The nitride NbN in the rocksalt structure is commonly reported as having the highest $T_{c}$ among the carbides and nitrides at 17.3 K.\cite{Toth, Williams} From a technological point of view, this material is interesting because of its possible application to induce superconductivity in carbon nanotube junctions.\cite{Kasumov:1999p91} Recent phonon calculations~\cite{Isaev:2007p7} showed that the rocksalt phase of NbN is unstable. It is known from experiment, however, that the rocksalt phase can be stabilized in the alloy NbC$_{1-x}$N$_{x}$ and in the nitrogen deficient NbN crystal.\cite{Chen:2005p21} In the present paper, we report \emph{ab initio} calculations of electron-phonon coupling in NbC$_{1-x}$N$_{x}$ and in the nitrogen deficient NbN to establish a parallel between the Kohn anomaly and $T_{c}$. We elaborate a model based on the Eliashberg spectral function, $\alpha^2F(\omega)$, explaining the enhancement of electron-phonon coupling due to the Kohn anomaly near the structure phase transition. 

\section{Method}

The calculations reported in the present study were carried out using density functional theory (DFT) within the local density approximation (LDA) as implemented in the ABINIT code.\cite{Gonze:2009p293} Norm-conserving Trouiller-Martins pseudopotentials were used to represent the interaction of the valence electrons with the atomic cores. The sampling of the Brillouin zone was done using a 24x24x24 $k$-grid with a gaussian broadening of the Fermi-Dirac distribution of 5 mHa. Wave functions of the electrons are expanded with a plane-wave basis up to an energy of 35 Ha. 

The NbC$_{1-x}$N$_{x}$ crystal was calculated within the virtual-crystal approximation (VCA) by simple mixing of the ionic pseudopotentials of carbon and nitrogen atoms.\cite{Ramer:2000p206, PhysRevB.62.R743} The nitrogen deficient NbN crystal was simulated by removing $x$ electrons per unit cell from a complete NbN crystal which we will denote by NbN$^{x}$. A uniform background charge maintains charge neutrality. Phonon spectra and electron-phonon coupling were evaluated within a linear response theory~\cite{Gonze:1997p223, Gonze:1997p222} on a 12x12x12 $q$-points sublattice. 

\section{Results}

The calculated density of states (DOS) is reported in Table~\ref{results}. In another study, Isaev et al. published results of \emph{ab initio} calculations for NbC.\cite{Isaev:2007p7} The main differences between their study and the present one are their use of ultrasoft pseudopotentials with the generalized gradient approximation (GGA) PBE for the exchange-correlation functional. We also used a denser grid of $k$-points for integration in the Brillouin zone (24x24x24 compared to 18x18x18) and a denser $q$-points grid for the phonon calculations (12x12x12 compared to 8x8x8). Our results agree with theirs to the expected precision of this type of approach given the different parameters used. 

The experimental data reported by Toth~\cite{Toth} are in agreement with our results given in Table~\ref{results}. In particular, the experimental value of the DOS at the Fermi level, $N_F$, for NbC is about 0.2~eV$^{-1}$ per atom, which corresponds to 0.4~eV$^{-1}$ per unit cell, which is within 10\% of our calculated value of 0.361~eV$^{-1}$ per unit cell. This agreement is satisfactory considering that the experimental value is obtained through specific heat measurements using the Sommerfeld free electron model. Furthermore, the deduced $\lambda$ from experimental data quoted by Toth employs the version of the McMilllan formula using the Debye temperature which is extracted again from the specific heat. Hence, a direct quantitative comparison with our results is not possible but we note that Toth quotes a value of ~0.6 for $\lambda$ whereas we obtain 0.682. Considering the differences mentioned above, this correspondence is acceptable and validates our approach.      

\begin{figure}
	\begin{center}
		\includegraphics[width=0.275\textwidth]{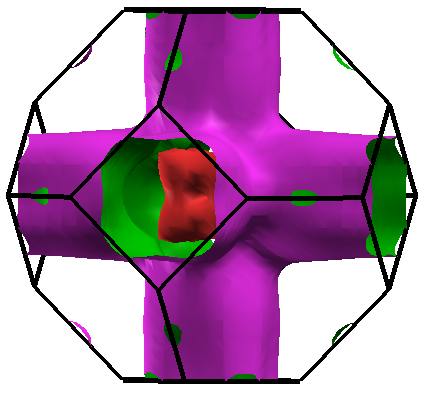}
	\end{center}
	\caption{\label{fig:nbc_FS}(color online). Fermi surface for NbC rendered by XCrysDen.\cite{Kokalj2003155} The cylinders exhibit important nesting for $\mathbf{k}$ in the $\Gamma X$ direction. Substituting carbon by nitrogen increases the radius of the cylinders.}
\end{figure}

\begin{figure} 
	\begin{center}
		\includegraphics[width=0.45\textwidth]{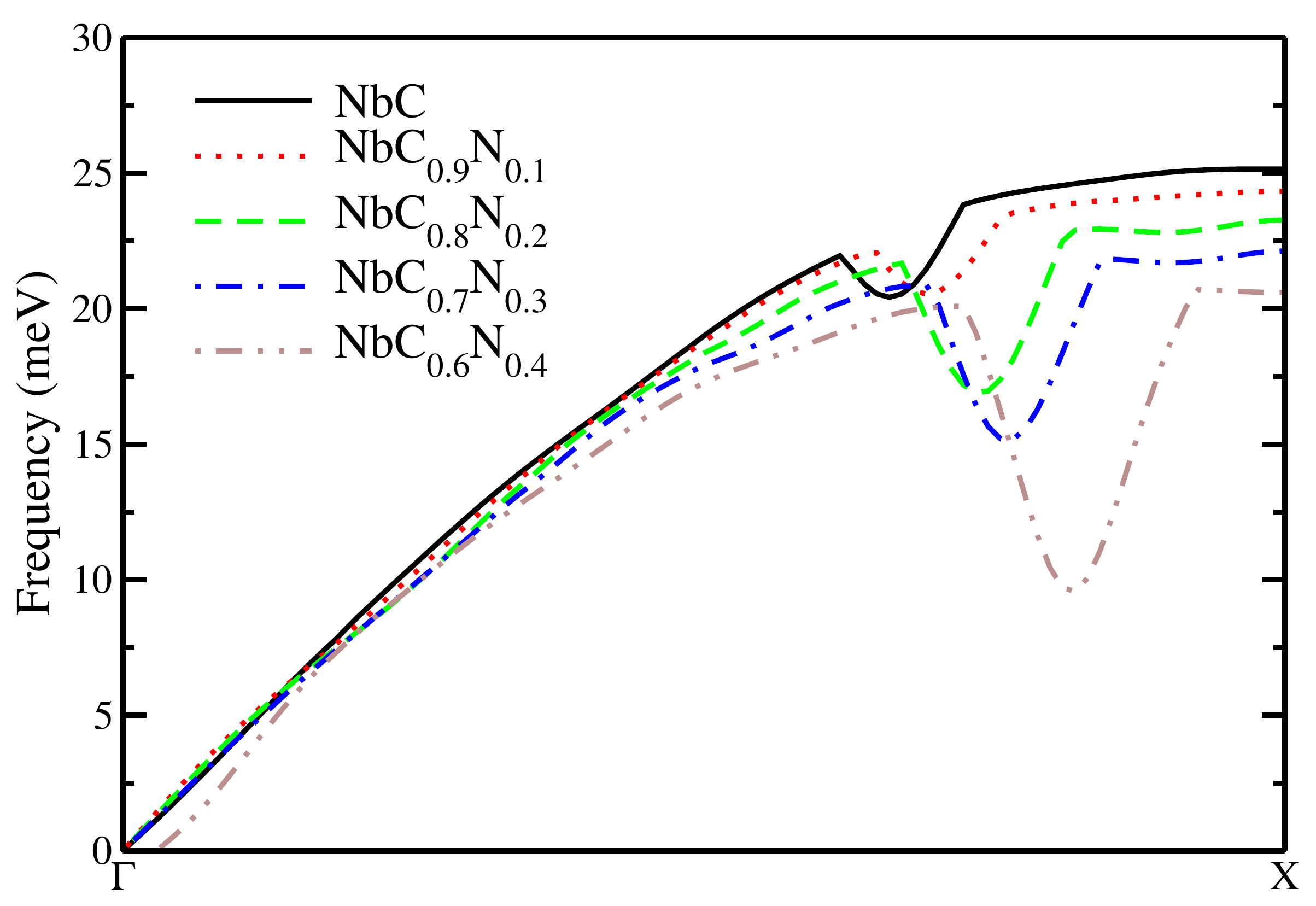}
	\end{center}
	\caption{\label{fig:phonon_gx}(color online). Phonon dispersion along the $\Gamma X$ direction for various concentration on $N$. The evolution of the Kohn anomaly is clearly seen. The softening becomes more important with the number of electrons and is also shifted towards $X$.}
\end{figure}

Upon inspection of the NbC Fermi surface, we see that it is made of arms lying along the $\Gamma X$ axis as depicted in Fig.~\ref{fig:nbc_FS}. This topology, which allows important nesting for $\mathbf{q}$ vectors connecting opposite faces of the arms, results in Kohn anomalies. This is similar to the case of TaC reported by Noffsinger et al.~\cite{Noffsinger:2008p93} For NbC$_{1-x}$N$_{x}$, adding electrons by partially substituting C by N atoms increases the radius of the arms of the Fermi surface which leads to enhanced nesting due to a larger phase space. This change of the Fermi surface shifts the wavevector of the resulting softened phonon towards larger $\mathbf{q}$. This phenomenon is readily observed in the phonon spectrum along $\Gamma X$ directions as shown in Fig.~\ref{fig:phonon_gx} for the longitudinal acoustic (LA) branch. 

\begin{figure} 
	\begin{center}
	\subfigure{\label{fig:phonon_nbc}}
		\includegraphics[width=0.45\textwidth]{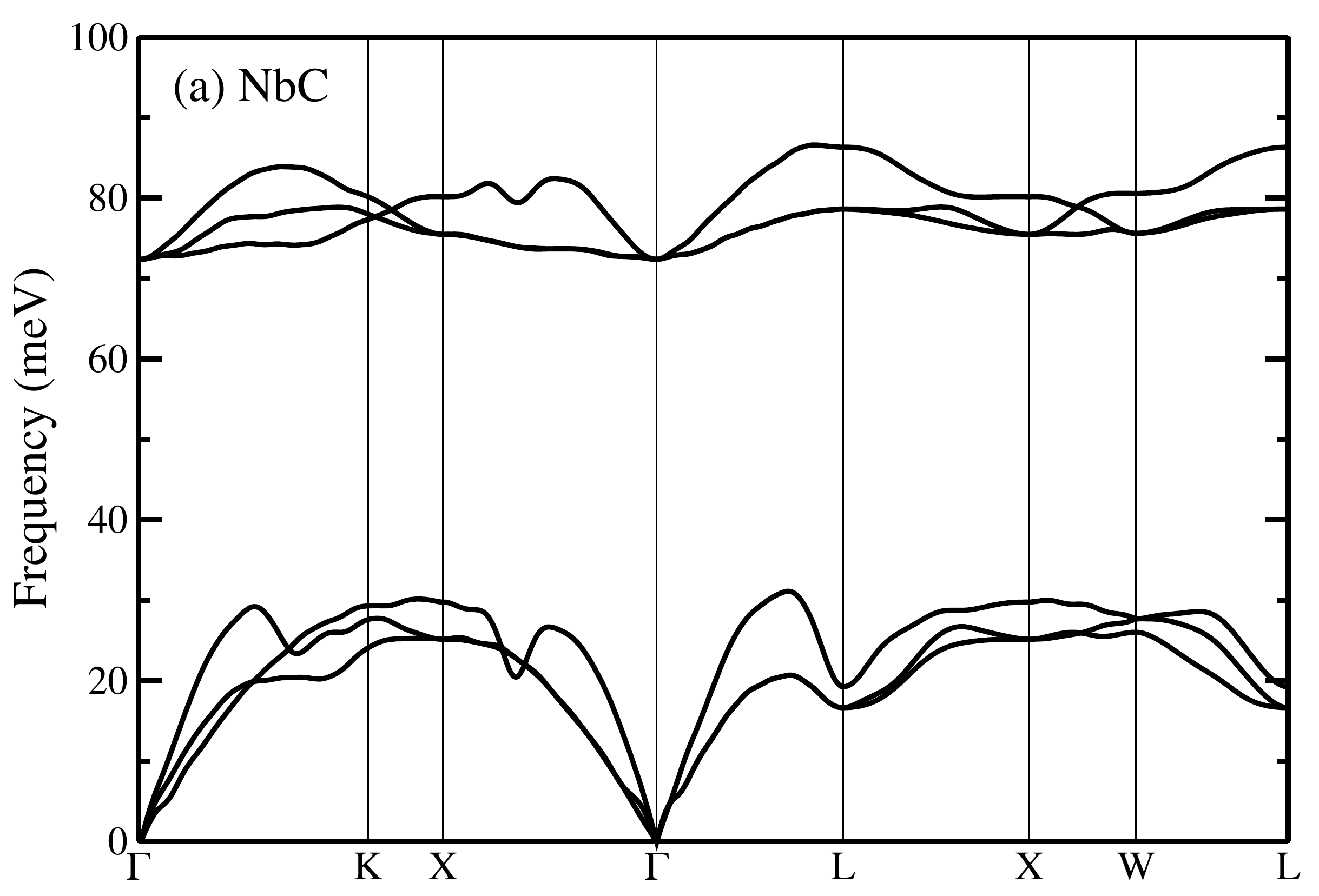}
	\subfigure{\label{fig:phonon_nbn}}
		\includegraphics[width=0.45\textwidth]{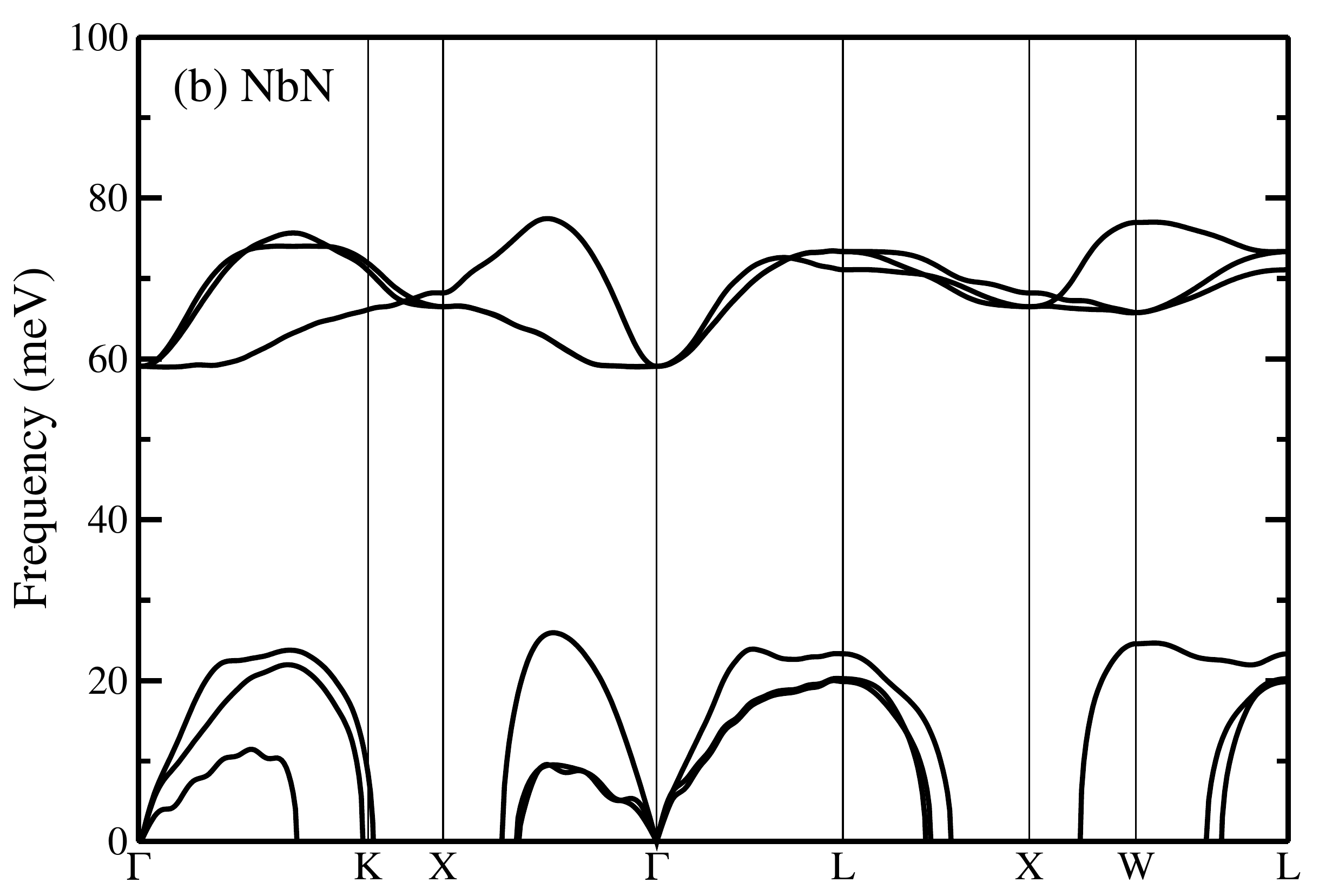}\\
	\vspace{.01 in}
	\subfigure{\label{fig:phonon_50N}}
		\includegraphics[width=0.45\textwidth]{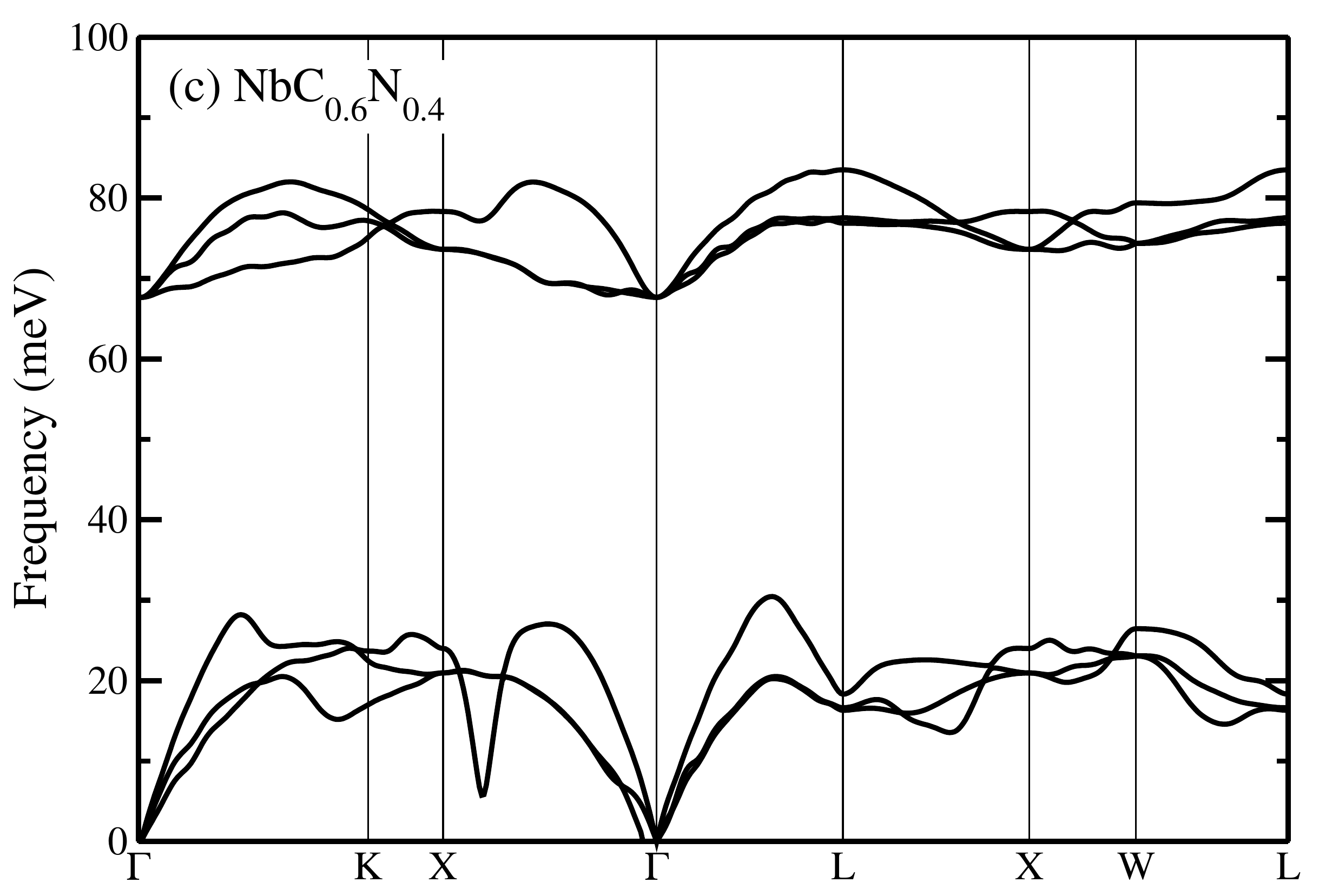}
	\subfigure{\label{fig:phonon_nbn04}}
		\includegraphics[width=0.45\textwidth]{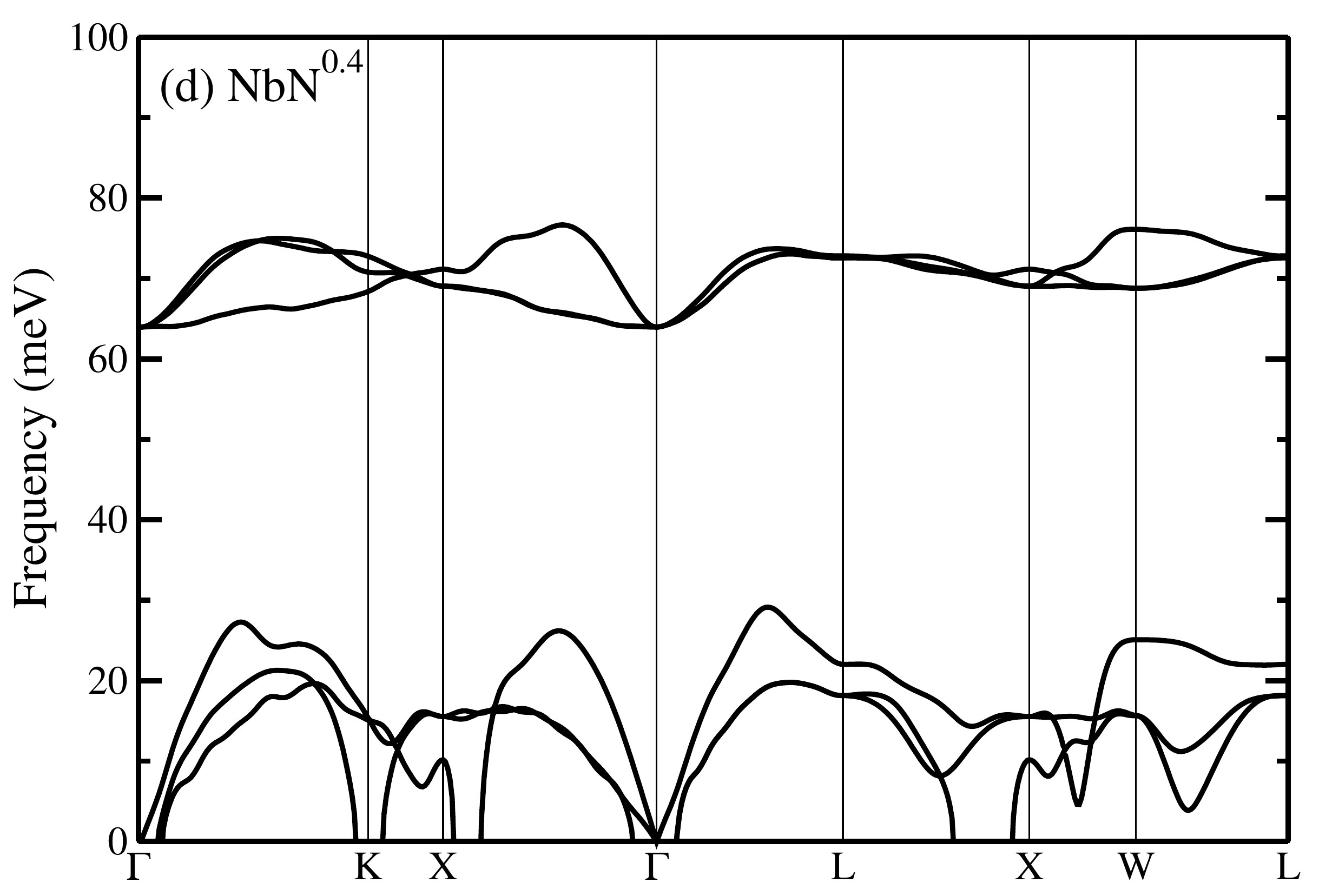}\\
	\vspace{.01 in}
	\end{center}
	\caption{Phonon band structure for (a) NbC, (b) NbN, (c) NbC$_{0.6}$N$_{0.4}$, (d) unstable NbN$^{0.4}$} \label{phonon} 
\end{figure}

The softening observed in the NbC spectrum, Fig.~\ref{fig:phonon_nbc}, is consistent with our analysis of the Fermi surface mentioned above. For NbC$_{0.5}$N$_{0.5}$ and NbN$^{0.4}$, the softening becomes an instability with imaginary phonon frequencies as illustrated in Fig.~\ref{fig:phonon_50N} and~\ref{fig:phonon_nbn04} respectively. We see from Fig.~\ref{fig:phonon_nbn} that NbN in the rocksalt structure is indeed unstable. We will refer to the density of states at the Fermi level of the structure at which the softening becomes an instability as $N_{c}$. The calculated values for this quantity are expected to be lower than the experimental values~\cite{Chen:2005p21} because we do not take into account anharmonic effects which are known to diminish the softening.\cite{Weber:1972p97} The quantity $N_{c}$ will prove to be fundamental in the elaboration of our theoretical model for the electron-phonon coupling. 

\begin{table*}[ht]
	\caption{\label{results}Density of states at the Fermi level, $N_{F}$, and results of linear response theory calculations for the electron-phonon coupling in NbC$_{1-x}$N$_{x}$ and in nitrogen deficient NbN. The total electron-phonon coupling constant, $\lambda$,  and the contribution of the acoustic branches, $\lambda_\textrm{ac}$, are reported separately. The weighted average of the phonon frequency, $\omega_\textrm{log}$, is needed to compute the transition temperature $T_{c}$ using the McMillan formula~\eqref{eq:McMillan}. The parameter $\mu^*$ is set to 0.1 in order to fit the NbC $T_{c}$ to its experimental value of 11.2~K.\cite{PhysRevB.8.5093} The experimental values of $T_c$ are also reported for comparison.\cite{Toth}}
	\begin{tabular}{ccccccc}   
	\hline
	$x$ &   ~~~$N_F$~~~                &~~~~~$\lambda$~~~~~&~~~~~$\lambda_\textrm{ac}$~~~~~&~~~$\omega_\textrm{log}$~~~&~~~$T_c$~~~&~~~Experimental $T_c$~~~\\
	    &(states eV$^{-1}$ cell$^{-1}$)&                   &                               &      (K)                  & (K)       & (K)\\
	\hline
	NbC$_{1-x}$N$_{x}$ & & & & & &\\
	\hline
	0 &  0.361 &  0.682 & 0.533 & 345 &  11.2 & 11.2 \\
	0.1 & 0.373 & 0.763 & 0.607 & 327 & 	14.3 &   12.5\\
	0.2 & 0.378 & 0.864 & 0.706 & 301  &  17.4 & 13.8 \\
	0.3 & 0.388 & 0.967 & 0.805 & 281  &  20.1 & 15.0\\
	0.4 & 0.405 & 1.108 & 0.943 & 260  &  23.1 & 16.2\\
	\hline
	NbN$^{x}$ & & & & & \\
	\hline
	1.0 & 0.360 &	0.516 &	0.367 &	373	 &	 7.1 &  \\
	0.9 & 0.375 & 0.624 & 0.462 & 348  &  8.9 & \\
	0.8 & 0.383 & 0.666 & 0.521 & 321  &  9.8 & \\
	0.7 & 0.398 & 0.818	& 0.661	& 297  &  14.6 & \\
	0.6 & 0.414 & 0.984	& 0.832	& 255  &  17.3 & \\
	0.5 & 0.420 & 1.943 & 1.782 & 145 &  20.4 & \\
	\hline
	\end{tabular}
\end{table*}

Table~\ref{results} reports the values, calculated within linear response theory, for the electron-phonon coupling constant $\lambda$, the weighted average of phonon frequencies $\omega_\textrm{log}$~\cite{Dynes} and the density of states at the Fermi level $N_{F}$. We have used the McMillan formula to estimate $T_{c}$~\cite{PhysRev.167.331}
\begin{equation}
    \label{eq:McMillan}
T_{c} = \frac{\omega_\textrm{log}}{1.20}\exp \left( \frac{-1.04(1+\lambda)}{\lambda-\mu^{*}(1+0.62\lambda)} \right)
\end{equation}
with $\mu^{*}=0.1$ to obtain an agreement between our calculations and the reported $T_{c}$ of $11.2~K$ for NbC.\cite{PhysRevB.8.5093} It is important to note that this equation is valid for $\lambda \leq 1$. In this context, the reported $T_c$ for NbN$^{0.5}$ is inaccurate as $\lambda$ is too large for the McMillan approximation.  Note that the McMillan formula can be used with different definition of the prefactor $\langle \omega \rangle$. We used the definition of $\omega_\textrm{log}=\exp \langle \ln \omega \rangle$ because it gives a better agreement between calculations and experiments~\cite{PhysRevB.12.905} but this has little effect on the model elaborated below. For NbC$_{1-x}$N$_x$, the rise of $T_c$ with the increase of the nitrogen concentration obtained in our calculations reproduces the trend in the experimental data~\cite{Toth}. The increase in the theoretical values is larger than the experimental one but the agreement is satisfactory considering the approximations involved in the McMillan equation. For NbN, the only experimental $T_c$ reported is 17.3~K.\cite{Toth, Williams} According to our study, the NbN in the rocksalt phase is only stable when the system is charged. We suspect that the experimental system is non-stoichiometric with deficient nitrogen. Therefore, it is appropriate to compare the experimental value with our NbN calculations near the structural instability. We see from Table~\ref{results} that we obtained a value of 20.4~K for a charge of 0.5 electron per unit cell, which compare well with the experimental value.

\section{Discussion}

The coupling constant $\lambda$ can be separated into two contributions, one from the optical modes and the other from the acoustic modes denoted $\lambda_\textrm{ac}$. The values of $\lambda_\textrm{ac}$ are reported in Table~\ref{results} together with $\lambda$. We can see that the variation of $\lambda$ as the density of states increases comes mainly from the change of $\lambda_\textrm{ac}$ while the optical part contributes only a constant independent of $N_F$. This is consistent with the hypothesis that the enhancement of the Kohn anomaly dominates the evolution in the electron-phonon coupling. Following this idea, we propose a simple model to express $\lambda$ and $T_{c}$ as functions of $N_{F}$. 

The values of $\lambda$ and $\omega_\textrm{log}$ can be extracted from the Eliashberg spectral function $\alpha^{2} F(\omega)$.~\cite{PhysRevB.6.2577} This function is defined as~\cite{Allen_Mitrovic}
\begin{align}
	\label{eq:a2f_def} 
& \alpha^2F (\omega)= \frac{1}{N_F^2} \sum_{kk'} \alpha^2F\left(k,k', \omega \right) \delta \left( \epsilon_k - \epsilon_F  \right) \delta \left( \epsilon_k' - \epsilon_F  \right) \\
& \alpha^2F (k,k',\omega) = \sum_j N_F |g^j_{k,k'}|^2 \delta \left( \omega - \omega_{\mathbf{k}-\mathbf{k'},j} \right) \nonumber \\
& g^j_{kk'} = \left( \hbar / 2M\omega_{\mathbf{k}-\mathbf{k'},j} \right)^{1/2} \langle k | \hat{\epsilon}_{\mathbf{k}-\mathbf{k'},j} \cdot \nabla V | k' \rangle \nonumber
\end{align}
where $\hat{\epsilon}_{\mathbf{q},j}$ is the normalized atomic displacement for a phonon. The sum over $j$ is over the different phonon branches. Since $g$ is inversely proportional to $\sqrt{\omega}$, the amplitude of the movement of the atoms is enhanced near the structural phase transition where $\omega$ becomes small for a soft mode, resulting in a stronger coupling. With all this in mind, we can represent the $\alpha^2 F(\omega)$ function by setting apart the contribution of the softened mode at frequency $\omega'$: 
\begin{equation}
	\label{eq:a2f}
\alpha^2 F(\omega) = N_{F} f_{0} (\omega) + \frac{|M|^{2} N_{F}}{\omega'} \delta \left(\omega-\omega'\right),
\end{equation}
\noindent
where $|M|^{2}$ represents an effective coupling matrix element of the relevant soft modes. From Green's function theory of electron-phonon coupling,\cite{Sadovskii} the renormalized phonon frequencies can be expressed in terms of the electron polarization, $\Pi(\omega, q)$ as 
\begin{equation}
\omega'^2=\omega_0^2 \left( 1 + \Pi(\omega, q) g^2 \right)
\end{equation}
where $g$ is the coupling and $\omega_0$ is the bare phonon frequency. Since the electron polarization is proportional to $-N_F$, the  renormalized phonon frequency of the form $\omega'=\omega_{0}\sqrt{1-N_{F}/N_{c}}$ is used to describe the evolution of the softened phonon, where $N_c$ is the critical density of states at which the structure becomes unstable.

The function $f_{0} (\omega)$ is independent of $N_{F}$ and represents the contribution from all the phonons that are not affected by the Kohn anomaly. Using~\eqref{eq:a2f}, we obtain an expression for $\lambda$:
\begin{equation}
	\label{eq:lambda}
\lambda= 2 \int d\omega \frac{\alpha^{2} F(\omega)}{\omega}= V_{0} N_{F} + \frac{2|M|^{2} N_{F}}{\omega'^{2}}
\end{equation}
where $V_{0} = 2\int d\omega f_{0}(\omega) \omega^{-1}$ is a material dependent parameter. This also gives a form for $\omega_\textrm{log}$
\begin{equation}
	\label{eq:wlog}
\omega_\textrm{log} = \exp \left[ \frac{2}{\lambda} \left( C N_{F}+\frac{|M|^{2} N_{F}}{\omega'^2} \ln \omega' \right) \right]
\end{equation}
where $C=\int d\omega f_{0}(\omega)\ln(\omega)\omega^{-1}$ is another parameter. Although this is a simple model for the Eliashberg spectral function, it allows a qualitative description of the effect of the soft phonon on the $T_c$. A possible improvement of the model would be to use a localized function in the second term of~\eqref{eq:a2f} to represent the phonon softening. For example, we could use a gaussian function to represent the softening which will allow to better describe the phonons affected. In order to simplify the model, a Delta function is used. 

We can express $\lambda (N_{F})$ by using only three parameters, $|M|^{2}\omega_{0}^{-2}$, $N_{c}$ and $V_{0}$. Our calculations of the phonon spectra yield an approximate value of $N_{c}$. We used a fit for the other parameters which are reported in Table~\ref{fit}. Results for these parameters for the case of NbC$_{1-x}$N$_{x}$ and the nitrogen deficient NbN are similar.

\begin{table}[hb]
\caption{Fit solution for the parameters in~\eqref{eq:lambda} and~\eqref{eq:wlog}. Contribution of the normal phonons to the electron-phonon coupling are represented by $V_{0}$. The softened frequency is taken into account by $|M|^{2}$ which is the coupling matrix element and by $\omega_{0}$, the bare frequency for these phonons. The critical density of states, $N_{c}$, indicates the occurrence of the structural phase transition. The parameter $C$ is obtained by inverting~\eqref{eq:wlog} and by using the values reported in Table~\ref{results}.} \label{fit}
\begin{tabular}{lcc}
	\hline
    ~ &   ~~NbC$_{1-x}$N$_{x}$    &    ~~NbN$^{x}$\\
	\hline
    $2|M|^{2} \omega_0^{-2}$ (eV cell) &	0.023 &	0.025 \\
	\hline
    $N_{c}$ (states eV$^{-1}$ cell$^{-1}$) &	0.414 & 0.424 \\
	\hline
    $V_{0}$ (eV cell) & 1.87 & 1.41\\
	\hline
	$C$ (eV cell $\log$(eV))  & -3.38 & -2.43 \\
	\hline
\end{tabular}
\end{table}

\begin{figure}
	\includegraphics[width=0.45\textwidth]{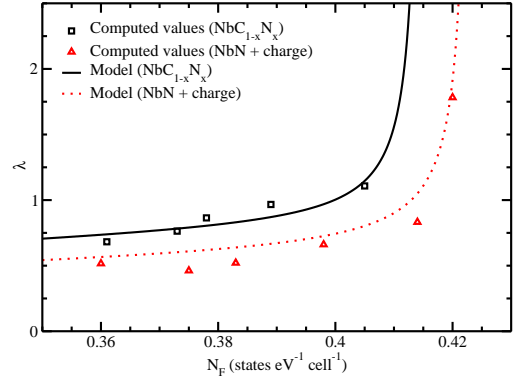}
	\caption{\label{graph_lambda}(color online). Fit solutions for $\lambda (N_{F})$ based on \eqref{eq:lambda} for mixed pseudopotential method (black curve) and for a uniform background charge (red curve). Parameters are fitted to the computed values and are reported in Table~\ref{fit}.}
\end{figure}

\begin{figure}
	\centering
	\includegraphics[width=0.45\textwidth]{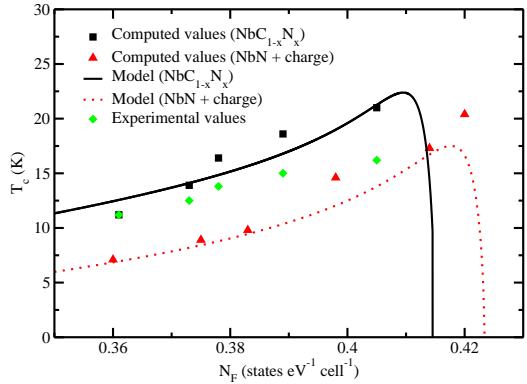}
	\caption{\label{graph:Tc}(color online). Best fit solutions for $T_{c} (N_{F})$ based on the McMillan equation~\eqref{eq:McMillan}. The $\omega_\textrm{log}$ term was computed using (\ref{eq:wlog}).   $\mu^{*}$ was fixed at 0.1 in order to get agreement between our results and the experimental $T_{c}$ for NbC.\cite{PhysRevB.8.5093} $T_{c}$ falls rapidly to zero near the critical DOS which is the onset of the structural phase transition. This is due to the inadequacy of the McMillan formula when $\lambda \geq 2$. The experimental values of $T_c$ reported on this figure are taken from~\cite{Toth} assuming that the relation between nitrogen concentration and the calculated DOS with alchemical pseudopotential is adequate.}  
\end{figure}

The model for $\lambda(N_{F})$ and the computed values are shown in Fig.~\ref{graph_lambda}. To compute $T_{c}$ using the McMillan formula~\eqref{eq:McMillan}, the bare phonon frequency is needed in order to determine $\omega_\textrm{log}$. Based on Fig.~\ref{fig:phonon_gx}, the renormalized frequency for the soft phonon is estimated at 20~meV for NbC. Hence, the bare frequency is estimated by $\omega_{0} = \omega'\left(1-N_{F}/N_{c}\right)^{-1/2}\approx 60$~meV. This gives an expression for $T_{c} (N_{F})$ which is plotted along with the calculated values in Fig.~\ref{graph:Tc}. Our model predicts an increase in $T_{c}$ as $N_{F}$ increases and a sharp fall when $N_{F} \approx N_{c}$. However, this feature appears in the region where $\lambda$ is very large. It is known that the McMillan equation is not adequate if $\lambda \geq 2$. In a recent work,\cite{moussa:094520} lower and upper bounds for $T_{c}$ have been derived:
\begin{equation}
f \left( \lambda \langle \omega^2 \rangle / \omega^2_\textrm{max}, \mu^* \right) \omega_\textrm{max} \leq T_c \leq f \left(\lambda, \mu^* \right) \sqrt{\langle \omega^2 \rangle}
\end{equation}
where
\begin{equation}
f\left( \lambda, \mu^* \right) = 0.69 \exp \left( -\frac{1+\lambda}{\lambda-\mu^*} \right) \sqrt{\frac{1+0.52\lambda}{1+5.6\mu^*}}
\end{equation}
is an empirically smooth function that interpolates between the weak-coupling limit for $T_c \propto \exp(-1/\lambda)$ and the strong-coupling one $T_c \propto \sqrt{\lambda}$. 

When used in our model, these bounds do not fall to zero at $N_{F} \approx N_{c}$ where $\lambda \rightarrow \infty$. Using the correct limit of $T_{c}$ for large $\lambda$~\cite{PhysRevB.12.905}:
\begin{equation}
	\label{eq:high_tc}
T_{c} \sim \sqrt{\lambda} \omega_{ph}
\end{equation}
where $\omega_{ph}$ is the average of phonon frequencies which can be taken to be $\omega_{\log}$. Near the phase transition, only the contribution of the soft frequency is important. Hence, if $N_{F} \approx N_{c}$, according to~\eqref{eq:lambda} and~\eqref{eq:wlog}, we find that $\lambda \sim \omega'^{-2}$ and $\omega_{\log} \sim \omega'$ so that~\eqref{eq:high_tc} gives $T_{c} \sim \sqrt{2|M|^{2} N_{c}}$ which is indeed not zero. 

\section{Conclusion}

In conclusion, we have presented \emph{ab initio} pseudopotential DFT results on NbC$_{1-x}$N$_{x}$ based on the virtual crystal approximation. We showed that the Kohn anomaly leads to an enhanced electron-phonon coupling. These results can be modeled by singling out the softening contribution to the Eliashberg spectral function.

\begin{acknowledgments}

This work was supported by grants from NSERC and FQRNT and by the NSF under Grant No. DMR07-05941 and the U.S. DOE under Contract No. DE-AC02-05CH11231. The computational resources were provided by the R\'eseau qu\'eb\'ecois de calcul de haute performance (RQCHP). We are grateful to Yann Pouillon for helpful and valuable technical support with the build system of ABINIT. 

\end{acknowledgments}

%

\end{document}